\documentclass[acmsmall]{acmart}
\AtBeginDocument{%
  }

\setcopyright{none}
\renewcommand\footnotetextcopyrightpermission[1]{}

\usepackage{enumitem}
\usepackage{xspace}
\usepackage{algorithm}
\usepackage{algorithmic}

\usepackage{graphicx}
\usepackage{listings}
\usepackage{xcolor}
\usepackage{enumitem}
\usepackage{amsmath}
\usepackage{float}
\usepackage{graphicx}
\usepackage{multirow}
\usepackage{caption}
\usepackage{subcaption}
\usepackage{xspace}
\usepackage{verbatim}
\usepackage{hyperref}
\usepackage{cleveref}
\usepackage{soul}
\usepackage{array}
\usepackage{adjustbox}
\usepackage{tabularx}

\usepackage[table]{xcolor}
\usepackage{wrapfig}
\usepackage{makecell}

\usepackage{pifont}
\usepackage{bbding}

\newcommand\rev[1]{\textcolor{black}{{#1}}}

\newcommand\tool{TestGeneralizer\xspace}

\newcommand\chg[1]{\textcolor{black}{{#1}}}

\usepackage{tikz}
\newcommand{\stepcircle}[1]{%
    \tikz[baseline=-0.7ex]{
    \node[circle, fill=black, text=white, font=\sffamily\scriptsize\bfseries,
      inner sep=0.35pt, minimum size=0.75em] (char) {#1};
  }%
}

\usepackage{varwidth}
\newcommand{\coloredbox}[2]{%
    \colorbox{#1}{\begin{varwidth}[t]{\dimexpr\linewidth-2\fboxsep\relax}#2\end{varwidth}}%
}

\definecolor{SoftRed}{RGB}{255, 228, 225}
\definecolor{SoftGreen}{RGB}{230, 255, 230}
\definecolor{SoftOrange}{RGB}{255, 229, 204}
\definecolor{SoftPurple}{RGB}{224, 224, 248}
\definecolor{SoftBlue}{RGB}{202, 240, 255}
\definecolor{SoftGray}{RGB}{237, 237, 237}
\definecolor{DeepGreen}{RGB}{0,100,0}
\definecolor{VioletCustom}{RGB}{138,43,226}

\begin{document}

\title{Generalizing Test Cases for Comprehensive Test Scenario Coverage}

\author{Binhang Qi}
\email{qibh@nus.edu.sg}
\affiliation{%
  \institution{National University of Singapore}
  \country{Singapore}
}

\author{Yun Lin}
\authornote{Corresponding author.}
\email{lin\_yun@sjtu.edu.cn}
\affiliation{%
  \institution{Shanghai Jiao Tong University}
  \country{China}
}

\author{Xinyi Weng}
\email{nanakusa@sjtu.edu.cn}
\affiliation{%
  \institution{Shanghai Jiao Tong University}
  \country{China}
}

\author{Chenyan Liu}
\email{chenyan@u.nus.edu}
\affiliation{%
  \institution{National University of Singapore}
  \country{Singapore}
}

\author{Hailong Sun}
\email{sunhl@buaa.edu.cn}
\affiliation{%
  \institution{Beihang University}
  \country{China}
}

\author{Gordon Fraser}
\email{Gordon.Fraser@uni-passau.de}
\affiliation{%
  \institution{University of Passau}
  \country{Germany}
}

\author{Jin Song Dong}
\email{dcsdjs@nus.edu.sg}
\affiliation{%
  \institution{National University of Singapore}
  \country{Singapore}
}

\begin{abstract}
Test cases are essential for software development and maintenance.  
In practice, developers derive multiple test cases from an implicit pattern based on their understanding of requirements and inference of diverse test scenarios, each validating a specific behavior of the focal method.  
However, producing comprehensive tests is time-consuming and error-prone: 
many important tests that should have accompanied the initial test are added only after a significant delay, sometimes only after bugs are triggered.

Existing automated test generation techniques largely focus on code coverage.  
Yet in real projects, practical tests are seldom driven by code coverage alone, since test scenarios do not necessarily align with control-flow branches.  
Instead, test scenarios originate from requirements, which are often \rev{undocumented and} implicitly embedded in a project's design and implementation.
\rev{However, developer-written tests are frequently treated as executable specifications; thus, even a single initial test that reflects the developer's intent can reveal the underlying requirement and the diverse scenarios that should be validated.}

In this work, we propose \tool, a framework for generalizing test cases to comprehensively cover test scenarios.  
\tool orchestrates three stages: 
(1) enhancing the understanding of the requirement and scenario behind the focal method and initial test; 
(2) generating a test scenario template and crystallizing it into various test scenario instances; and 
(3) generating and refining executable test cases from these instances.  
To ensure accuracy and completeness, \tool combines rule-based prompts, automatically optimized via a prompt auto-tuning technique, with crucial project knowledge retrieved through program analysis.
We evaluate \tool against three state-of-the-art baselines (EvoSuite, gpt-o4-mini, and ChatTester) on 12 open-source Java projects, covering 506 multi-test focal methods and 1,637 test scenarios.  
\tool achieves significant improvements: \rev{+57.67\% and +59.62\% over EvoSuite, +37.44\% and +32.82\% over gpt-o4-mini, and +31.66\% and +23.08\%} over ChatTester, in mutation-based and LLM-assessed scenario coverage, respectively.  
In a field study, we submitted 27 generalized tests overlooked by developers; 16 were accepted and merged into official repositories, demonstrating the practical usefulness of \tool.
\end{abstract}



\maketitle

\section{Introduction}

Software testing plays a crucial role in ensuring the quality of both open-source and industrial software by verifying whether implementations satisfy user requirements.  
A large body of research has been devoted to automating test generation, which has been predominantly \textbf{code coverage-driven}.
Classical approaches~\cite{cadar2008klee,godefroid2005dart,pacheco2007randoop,lin2021graph} such as EvoSuite~\cite{fraser2011evosuite} frame test generation as the problem of maximizing structural coverage (e.g., branches or paths).  
This formulation reduces test generation to a constraint-solving problem, giving rise to techniques including symbolic execution~\cite{cadar2008klee,braione2017combining,braione2018sushi,godefroid2005dart,sen2005cute} and search-based testing~\cite{fraser2011evosuite,arcuri2008search,braione2017combining,godoy2021enabledness,lin2021graph,lemieux2023codamosa,pacheco2007randoop,lin2020recovering}.
With the rise of large language models (LLMs),
recent work~\cite{nie2023learning,tufano2020unit,kang2023large,dinella2022toga,chattester,IntUT,lemieux2023codamosa} recasts test generation as a special case of code generation. 
Given a focal method and an instruction prompt, these approaches leverage LLMs (e.g., ChatGPT) to inductively produce test code by translating the focal method, the prompt, or both into executable test cases.
Yet despite their different form, these LLM-based approaches remain code coverage-driven: their primary goal is to improve code coverage, whether by optimizing prompts and pipelines~\cite{chattester,IntUT} or by integrating LLMs with classical techniques~\cite{lemieux2023codamosa,TELPA}.

While existing test generation approaches have shown promising results,
practical test cases are rarely written solely to maximize coverage metrics or exercise focal code.
In practice, developers design test cases around diverse \textbf{test scenarios} that validate whether a focal method satisfies specific requirements.
A test scenario is derived from a requirement, aiming to validate whether a specific behavior of a focal method is expected by the requirement. 
For example, in the \texttt{ofdrw} project~\cite{ofdrw}, the requirement (obtained by comments) for the focal method \texttt{setPaint(Paint paint)} is:
``To generate OFD through graphic drawing, it must be possible to set the brush color for both filling and stroking.''
To validate this requirement, developers wrote multiple tests to cover various test scenarios, including setting the color as a solid, a multi-stop linear gradient, and a radial gradient.
Designing high-quality tests to comprehensively cover such scenarios is important, 
yet time-consuming and error-prone.
It is often the case that developers realize important test scenarios only later, sometimes after failures occur.
In this example, the test \texttt{linearGradientPaint()} was first added in commit \texttt{91af2eb}~\cite{ofdrwCommit1} (February 1, 2023), while the test \texttt{setPaintRadialGradientPaint()} was only introduced in commit \texttt{9f82f37}~\cite{ofdrwCommit2} (March 6, 2023).
Notably, we verified that the latter test also executes successfully on the earlier version, indicating that it should have been included from the initial commit.
A similar case arises in the \texttt{cron-utils} project, where a developer supplements an additional test case~\cite{cronUtilsTest} to validate an overlooked test scenario: ``infinite loop when daylight savings time starts at midnight,'' raised through an issue report~\cite{cronUtilsIssue}.
These examples highlight the importance and the difficulty of deriving comprehensive test scenarios.
This motivates the following research question:
\textit{Given a focal method and a developed test case, how can we generalize it into additional test cases that comprehensively cover the test scenarios intended by the developer?}

Apparently, code coverage-driven solutions, whether classical~\cite{cadar2008klee,godefroid2005dart,pacheco2007randoop,lin2021graph} or LLM-based~\cite{nie2023learning, tufano2020unit, kang2023large, dinella2022toga, chattester,lemieux2023codamosa,IntUT,TELPA}, are not well-suited for this task, 
since test scenarios do not necessarily align with control-flow branches
(e.g., \texttt{setPaint()} contains no branches).
The key challenge lies in \textbf{Implicit Functionality Requirement}: 
complete and well-maintained requirements are rare, especially under agile development.
Instead, requirements are often only implicitly embedded in the project’s functional design and developer intent.
Without understanding such implicit requirements, it is difficult to infer comprehensive test scenarios without omissions or redundancies.
\rev{This setting also differs from established paradigms like property-based testing~\cite{proze,maciver2019hypothesis,vikram2023can,tiwari2021production}, which rely on explicit properties.
In practice, behavioral intent is often implicitly encoded in developer-written tests.
We hypothesize that such tests contain a latent scenario template, and that a single initial test can provide sufficient information to be generalized into diverse, behaviorally consistent tests.}

In this work, we propose \tool, a three-stage framework for generalizing test cases to cover comprehensive test scenarios.
Across these stages, \tool addresses the implicit requirement challenge through \textit{project knowledge collection} and \textit{prompt auto-tuning}, enabling accurate generalization from a single test case.
Specifically, given a focal method and one initial test case, \tool proceeds as follows.
In Stage \stepcircle{1} \rev{(Enhancing Test Scenario Understanding)}, \tool transforms the initial test into multiple-choice ``exams'' by mutating its oracles.
The LLM is asked to select the correct oracle, providing a measurable check of its understanding of the requirement and the current test scenario.
In Stage \stepcircle{2} \rev{(Test Scenario Generalization)}, \tool prompts the LLM to generate a \textit{test scenario template}, a semi-structured plan describing how to validate the focal method under diverse test scenarios.
The template consists of step-by-step actions with \textit{variation points} (VPs), whose different settings derive distinct intended test scenarios.
By fixing the VP settings, the template is crystallized into concrete \textit{test scenario instances}, each validating a meaningful behavior of the focal method.
Each instance contains both \textit{primary oracles} (deduced from project design and implementation) and \textit{alternative oracles} (inferred from common requirement knowledge).
In practice, alternative oracles can alert developers to potential design flaws or implementation bugs.
In Stage \stepcircle{3} \rev{(Test Generation)}, each test scenario instance serves as guidance for the LLM to generate a concrete test. 
Generated tests are iteratively refined with the help of project knowledge, resolving compilation errors, execution errors, and assertion failures until the test passes or a maximum iteration limit is reached.
\rev{To make this pipeline effective, \tool overcomes two primary challenges. \textbf{(1) Implicit Pattern Recognition}: distinguishing true scenario VPs from noisy code elements is highly complex. \tool addresses this through a prompt auto-tuning technique (in Stage \stepcircle{2}) that automatically learns precise rules for VP identification. \textbf{(2) Reasonable Scenario Crystallization}: determining semantically valid and project-specific settings for these VPs is error-prone. \tool overcomes this by proactively retrieving crucial project knowledge (in Stage \stepcircle{1} --- \stepcircle{3}) via program analysis, grounding generalized scenarios in the actual project context rather than relying on LLM hallucination.}

We extensively evaluate \tool on 506 focal methods from 12 open-source projects, involving 1,637 test scenarios.
Compared to state-of-the-art approaches (e.g., gpt-o4-mini~\cite{gpt-o4-mini}, ChatTester~\cite{chattester} and EvoSuite~\cite{fraser2011evosuite}),
\tool demonstrates clear advantages.
In particular, compared to ChatTester, \tool achieves:
\textbf{(1) Higher Scenario Coverage:}
\tool generalizes more comprehensive test scenarios, improving scenario coverage by \rev{31.66\% and 23.08\%} in mutation-based (i.e., overlap between mutants killed by generated and ground-truth tests) and LLM-assessed scenario coverages, respectively. 
Also, the effectiveness of \tool is consistent on both commercial LLMs (e.g., ChatGPT) and open-source LLMs (e.g., DeepSeek-V3.1).
\textbf{(2) Practical Impact:}
In a field study, we submitted pull requests with tests generalized by \tool but overlooked by developers. Of the 27 submitted tests, 16 were accepted and merged into official repositories, demonstrating the practical usefulness of \tool.

In summary, this work makes the following contributions:
\begin{itemize}[leftmargin=*]
  \item \textbf{Methodology.} To the best of our knowledge, \tool is the first approach to generalize test cases.
  We introduce techniques for enhancing requirement understanding and inferring test scenarios,
  laying the groundwork for deriving tests directly from requirements.
  \item \textbf{Dataset and Evaluation.}
  We release a benchmark for test generalization, including focal code, test code, and high-quality test scenario templates.
  The benchmark consists of 506 multi-test focal methods from 12 open-source projects, involving 1,637 test scenarios, curated to ensure quality and practice relevance.
  Using this benchmark, we extensively evaluate \tool against state-of-the-art baselines.
  Results show that \tool generalizes more comprehensive test scenarios and consistently outperforms baselines in scenario coverage.
  \item \textbf{Field Study.}
  To assess practical usefulness, we conduct a field study by submitting pull requests with test cases generalized by \tool but overlooked by project developers.
  Of the 27 submitted tests, 16 were accepted and merged into official repositories.
\end{itemize}

All source code, benchmark, experimental results, and field study data are available at \url{https://github.com/code-philia/TestGeneralizer}.

\lstset{
  language=Java,
  basicstyle=\ttfamily\fontsize{6pt}{6pt}\selectfont,
  keywordstyle=\color{blue},
  stringstyle=\color{VioletCustom},
  commentstyle=\color{blue},
  numbers=left,
  numberstyle=\tiny\color{black},
  stepnumber=1,
  numbersep=5pt,
  showspaces=false,
  showstringspaces=false,
  tabsize=4,
  breaklines=true,
  breakatwhitespace=false,
  showtabs=false,
  frame=single,
  linewidth=0.98\columnwidth,
  escapeinside={(*@}{@*)},
  captionpos=b,
  belowcaptionskip=-10pt,
  xleftmargin=10pt
}

\section{Motivation Example}
\label{sec:motivation}

In practice, when developers write test cases to validate a requirement, they often follow a recurring pattern to crystallize multiple test scenarios from this pattern and derive corresponding test cases.

\setlength{\intextsep}{3pt}
\begin{wrapfigure}[5]{r}{0.42\textwidth}
  \centering
  \captionsetup{skip=4pt}
  
  \includegraphics[width=0.42\textwidth]{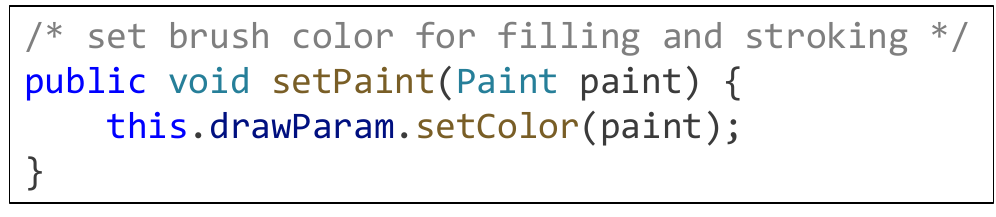}
  \caption{Focal method in \texttt{ofdrw} project~\cite{ofdrw}.}
  \label{fig:motivation-1-fm}
  \vspace{-10pt}
\end{wrapfigure}

\begin{figure*}[t]
  \centering
  \captionsetup{skip=4pt}
  \includegraphics[width=1\textwidth]{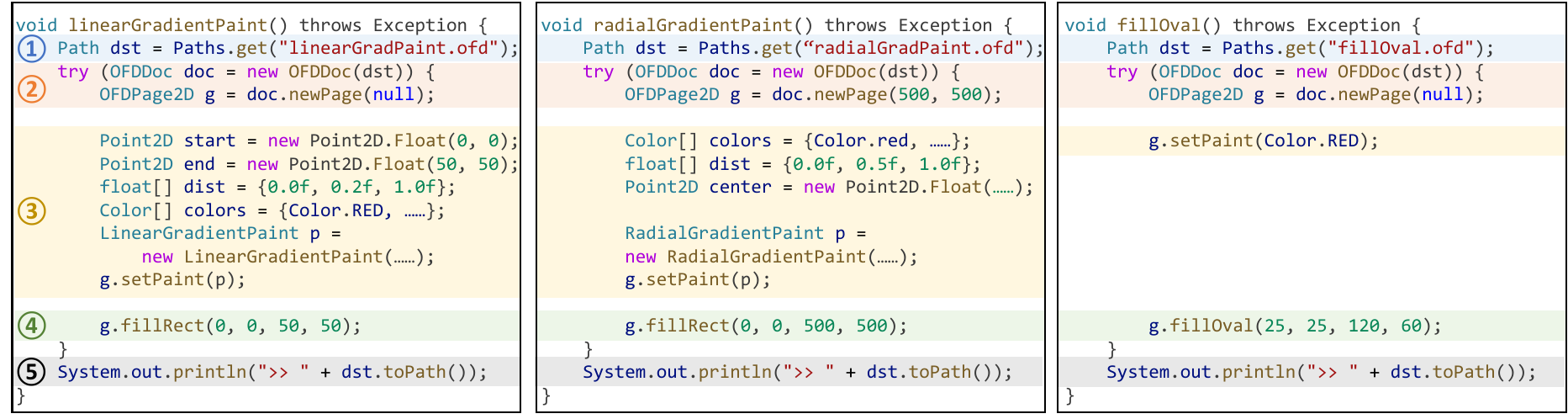}
  \caption{Test cases for the focal method \texttt{setPaint()}, covering three test scenarios. 
  The tests follow a common pattern consisting of five parts (highlighted by colored backgrounds), which collectively validate the requirement behind the focal method.}
  \vspace{-8pt}
  
  \label{fig:motivation-1-tests}
\end{figure*}

For example, \autoref{fig:motivation-1-fm} shows a focal method in \texttt{ofdrw}~\cite{ofdrw}, which allows users to set the brush color for both filling and stroking when generating OFD through graphic drawing.
\autoref{fig:motivation-1-tests} presents the corresponding tests written by developers to validate this requirement.
By comparing these tests, we observe that they follow a shared pattern consisting of five parts, highlighted in the figure.
If such a pattern is already identified and crucial project knowledge (e.g., \texttt{LinearGradientPaint}, \texttt{RadialGradientPaint}, \texttt{fillRect}, and \texttt{fillOval}) is already known, generalizing the test scenarios and implementing the tests is straightforward.
However, when only a single test case is available as a reference, generalizing the remaining test scenarios and writing their corresponding tests becomes non-trivial.
This challenge arises for several reasons:

\begin{itemize}[leftmargin=*]
  \item \textbf{Implicit Pattern Recognition.}
  Focal methods often provide little guidance on which test scenarios are intended, since scenarios do not necessarily correspond to control-flow branches and branchless methods such as \texttt{setPaint()} are common.
  Even with an initial test as a reference, it is still difficult to infer the underlying pattern and variation points (VPs) regarding test scenarios from numerous changeable code elements.
  For example, with \texttt{linearGradientPaint()} as the initial test, one can observe many changeable code elements across the five parts, such as the path in part \ding{172}, the parameter of \texttt{newPage()} in part  \ding{173}, the parameters of \texttt{LinearGradientPaint()} and \texttt{setPaint()} in part \ding{174}, and the filling method and its parameters in part \ding{175}.
  Yet test scenarios involve three VPs: the canvas setting, the paint style, and the drawing shape.
  These correspond to only three code elements, including 
  the parameters of \texttt{newPage()} and 
  \texttt{setPaint()}, 
  and the choice of filling method.
  Other code elements are either trivial (e.g., the parameters of \texttt{fillRect()}) or dependent on these VPs (e.g., the path).
  Thus, inferring the underlying pattern with intended VPs is highly challenging.
  
  \item \textbf{Reasonable Scenario Crystallization.}
  Even having accurate VPs, determining which test scenarios are reasonable to crystallize remains difficult.
  Feasible choices for VP settings may be unknown, or syntactically valid but semantically unreasonable with respect to requirements.
  For example, deriving these tests based on the pattern requires several project-specific knowledge:
  \texttt{fillOval()} is an alternative to \texttt{fillRect()} with similar functionality, while \texttt{RadialGradientPa\\int()} and \texttt{Color} are feasible parameters for \texttt{setPaint()}.
  Without such knowledge, guessing appropriate APIs or parameter values is difficult and prone to error.
\end{itemize}

Facing these challenges, code coverage-driven test generation often produces redundant test cases and misses important scenarios, leading to inefficiency and incomplete validation. 

\noindent\textbf{Coverage-driven Test Generators}.
Traditional coverage-driven approaches rely solely on control-flow branches, without awareness of requirements or test scenarios.
Since test scenarios do not necessarily correspond to branches, these approaches cannot capture the intended behaviors of the focal method.
In this example, coverage-driven software testing tools such as EvoSuite~\cite{fraser2011evosuite} stop immediately after generating a single executable test case.

\begin{lstlisting}[float=t, caption={ChatTester-generated tests, with one missing and three redundant cases relative to ground truth.}, label=fig:motivation-1-chattester]
void testSetPaintWithBasicColor() {
    DummyOFDPageGraphics2D g = new DummyOFDPageGraphics2D();
    Color testColor = Color.BLUE;
    g.setPaint(testColor);
    assertEquals(testColor, g.getPaint(), "The paint should be updated to Color.BLUE");
}

void testChangingPaintMultipleTimes() {  (*@\coloredbox{SoftRed}{\ding{55} \textbf{Redundant test scenario}}@*)
    DummyOFDPageGraphics2D g = new DummyOFDPageGraphics2D();
    Color firstColor = Color.MAGENTA;
    g.setPaint(firstColor);
    assertEquals(firstColor, g.getPaint(), "Paint should be updated to Color.MAGENTA");
    // ...
    LinearGradientPaint secondPaint = new LinearGradientPaint(start, end, dist, colors);
    g.setPaint(secondPaint);
    assertEquals(secondPaint, g.getPaint(), "Paint should now be updated to the new GradientPaint");
}

void testSetPaintWithNull() { (*@\coloredbox{SoftRed}{\ding{55} \textbf{Redundant test scenario}}@*)
    DummyOFDPageGraphics2D g = new DummyOFDPageGraphics2D();
    g.setPaint(null);
    assertEquals(null, g.getPaint(), "Paint should be null after setting it to null");
}

public void linearGradientPaintWithTransform() throws Exception {  (*@\coloredbox{SoftRed}{\ding{55} \textbf{Redundant test scenario}}@*)
    final Path dst = Paths.get("target/linearGradientPaintWithTransform.ofd");
    try (OFDGraphicsDocument doc = new OFDGraphicsDocument(dst)) {
        OFDPageGraphics2D g = doc.newPage(500, 500);
        g.translate(250, 250);
        g.rotate(Math.toRadians(30));
        g.scale(1.2, 1.2);
        // ...
        LinearGradientPaint p = new LinearGradientPaint(start, end, dist, colors);
        g.setPaint(p);
    }
    System.out.println(">> " + dst.toAbsolutePath());
}
\end{lstlisting}

\noindent\textbf{LLM-based Test Generators}.
\autoref{fig:motivation-1-chattester} shows the results of ChatTester~\cite{chattester}, a state-of-the-art test generation solution built on GPT-o4-mini.
We adapt ChatTester for test generalization by additionally providing it with an initial test (i.e., \texttt{linearGradientPaint}) and modifying its prompt from ``generate one test case'' to ``generate more test cases with reference to the initial test case,'' while keeping all other settings as default.
As shown in \autoref{fig:motivation-1-chattester}, 
the generalized tests produced by ChatTester remain considerably distant from the ground-truth tests, due to two major issues:
\textbf{(1) Failure to capture the underlying pattern.}
The generated tests miss key parts of the pattern and the associated VPs.
For instance, \texttt{testSetPaintWithBasicColor()} misses the first part of setting a path and the final part of invoking a filling method.
\textbf{(2) Failure to infer reasonable test scenarios.}
While ChatTester does infer a scenario involving \texttt{Color} (corresponding to the ground-truth test \texttt{fillOval()}), likely guided by common knowledge and hints in the initial test (where \texttt{Color} is already used), it fails to produce the important scenario \texttt{radialGradientPaint()}.

To address these challenges, we introduce \tool, a framework for accurately generalizing comprehensive test scenarios.
\tool first constructs a test scenario template with precise VPs, guided by a high-quality prompt with well-designed rules obtained through prompt auto-tuning.
It then retrieves crucial project knowledge to strengthen the understanding of requirements and test scenarios, enabling the crystallization of reasonable test scenario instances from the template.
From these instances, it generates corresponding test cases, achieving comprehensive coverage of the test scenarios targeted by ground-truth tests.
Moreover, \tool produces additional tests beyond the ground-truth; several of these were submitted as pull requests and successfully merged by project maintainers, as detailed in our field study (Section \ref{sec:field_study}).

\section{Approach}
\begin{figure*}[t]
  \centering
  \includegraphics[width=\textwidth]{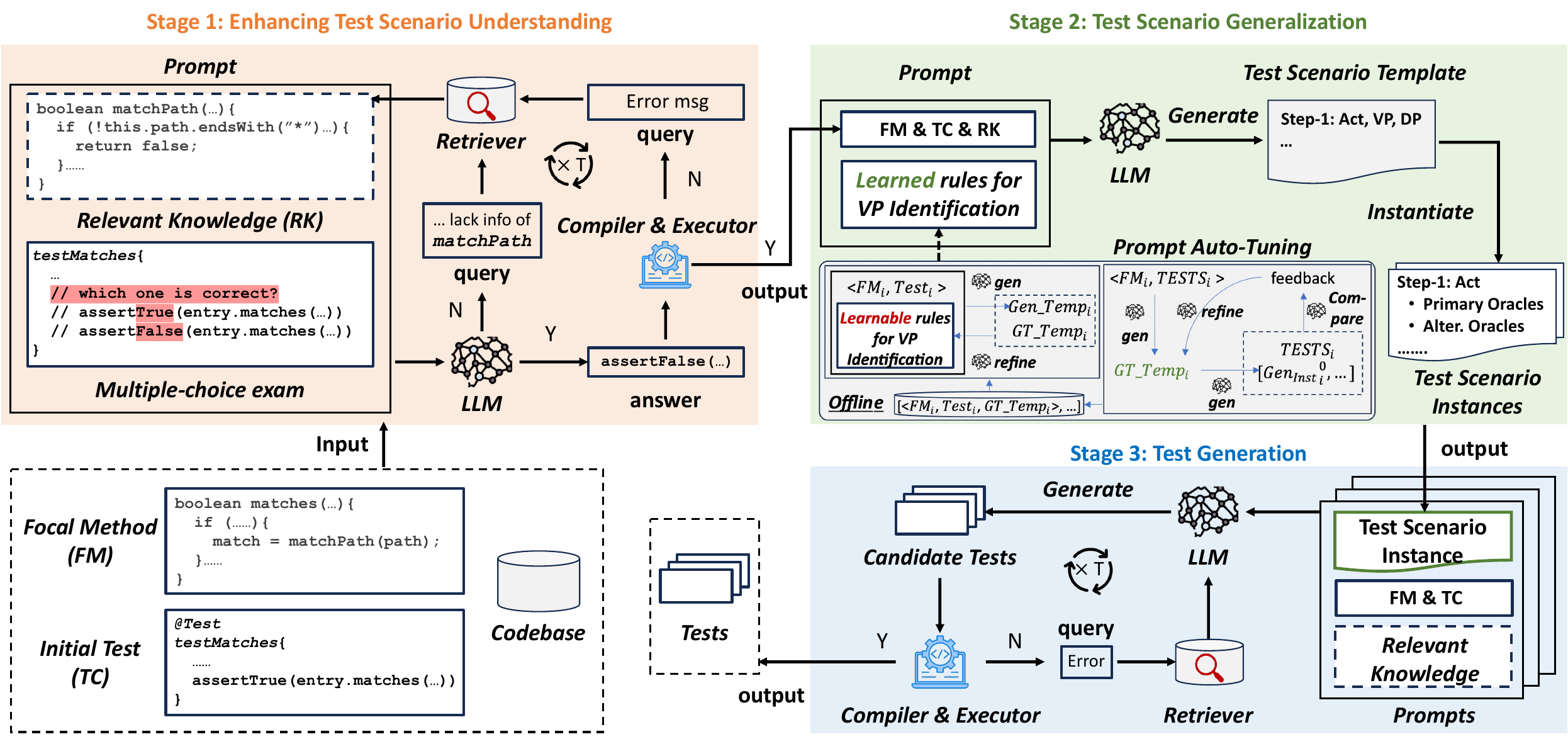}
  \caption{Overview of \tool: Given a focal method, an initial test, and the codebase, \tool generalizes the initial test into additional tests that comprehensively cover intended scenarios. \chg{The retriever is implemented using predefined CodeQL and JDTLS script templates instantiated with LLM-generated queries.}}
  \vspace{-8pt}
  \label{fig:overview}
\end{figure*}

\begin{table}[]
\caption{\chg{Definitions of core terminology used in \tool.}}
\vspace{-6pt}
\label{tab:term}
\footnotesize
\begin{tabularx}{\linewidth}{cX}
\toprule
\multicolumn{1}{c}{\textbf{Term}} & \multicolumn{1}{c}{\textbf{Definition}}                                                                                                                                                                                                                      \\ \midrule
\makecell[tc]{Relevant\\project knowledge}        & Project-specific facts retrieved from the codebase, such as method bodies, class relationships, and field definitions, that help understand requirements, identify variation points, determine feasible settings, or repair generated tests. \\ \midrule
\makecell[tc]{Variation point\\(VP)}                   & A scenario-relevant factor, either a code element or high-level testing   choice, whose alternative settings produce distinct intended behaviors of the   focal method                                                                                       \\ \midrule
\makecell[tc]{Test scenario\\template}            & A general, semi-structured test plan that abstracts the developer’s intended testing pattern into steps, variation points, and dependencies among steps.                                                                                                 \\ \midrule
\makecell[tc]{Test scenario\\instance}            & A concrete instantiation of a test scenario template, obtained by fixing all variation-point settings and determining the corresponding oracle(s).  \\ \bottomrule
\end{tabularx}
\end{table}

\autoref{fig:overview} illustrates the workflow of \tool, which consists of the following three stages.
\chg{\autoref{tab:term} summarizes the core terminology used throughout
the approach.}
{
\color{black}
\begin{itemize}[leftmargin=*]
  \item \textbf{Stage 1 (Enhancing Test Scenario Understanding):}
  Given a focal method $fm$ and a developer-written test $tc$, \tool infers the underlying requirement and test scenario reflected by $tc$.
  \chg{To achieve this, \tool introduces \textit{Masked Oracle Modeling} (MOM), which validates and enhances the LLM's understanding of implicit requirements and test scenarios. MOM asks the LLM to recover the original oracle from executable wrong-oracle alternatives, and triggers retrieval of relevant project knowledge from $K$ when the LLM is uncertain or incorrect.}
  
  \item \textbf{Stage 2 (Test Scenario Generalization):}
  Building on the understanding from Stage 1, \tool produces a \textit{test scenario template} $so_{temp}$, a concise step-by-step plan describing how to test $fm$ and what variation points (VPs) are.
  The template is then crystallized into a set of \textit{test scenario instances} $\{so_{ins}^m\}_{m=1}^M$, each with concrete VP settings and corresponding oracles.
  To improve the comprehensiveness of $\{so_{ins}^m\}_{m=1}^M$, 
  \tool (1) prompts the LLM to proactively query relevant knowledge $K$ to recover potential VPs absent from the current context, and (2) applies prompt auto-tuning to optimize VP identification.
  
  \item \textbf{Stage 3 (Test Generation):}
  Finally, \tool generates executable tests from $\{so_{ins}^m\}_{m=1}^M$.
  For each instance, 
  the LLM is given $fm$, $tc$, $K$, and $so_{ins}^m$ to produce a test $tc_{gen}$.
  It iteratively repairs failing tests using compiler and runtime feedback together with retrieved knowledge.
\end{itemize}
}

\subsection{Enhancing Test Scenario Understanding}
Understanding the underlying requirements and test scenarios is the foundation of test generalization. 
Since requirements and scenarios are often implicitly embedded in project design, collecting relevant project knowledge is essential for improving the understanding.
However, even with a developer-written test for reference, 
two challenges remain: 
(1) determining whether the current context already contains sufficient knowledge, and 
(2) identifying which pieces of knowledge are important amid the abundance of potentially noisy information.
Relying solely on the LLM to judge sufficiency is unreliable due to hallucination.
To address this, we propose \chg{\textit{Masked Oracle Modeling} (MOM), which examines the LLM using multiple-choice exams that ask it to recover the original oracle from executable wrong-oracle alternatives.}
These exams provide a more deterministic measure of the LLM's understanding while also guiding the retrieval of relevant knowledge.

\noindent\textbf{Exam Generation.}
Given $fm$ and $tc$ with assertions, \tool prompts the LLM to generate up to $Q$ incorrect oracles (e.g., $Q=10$) for each assertion in $tc$.
The key instruction is: \textit{``1. Identify the assertions and oracles used in the test case. 2. Come up with up to 10 WRONG oracles for each identified oracle. NOTE: The oracle is wrong, but the assertion statement should still be compilable, which means you cannot invent APIs or fields out of thin air.''}
The LLM generates wrong oracles by, for example, altering expected results or modifying parameter values in the invocation of $fm$.  
As shown in Stage 1 of \autoref{fig:overview}, one wrong oracle changes \texttt{assertTrue} to \texttt{assertFalse}.
\tool filters out wrong oracles that cause compilation or execution errors, retaining only those that trigger assertion failures.
If no valid wrong oracles remain, the LLM is instructed to revise its output. 

\noindent\textbf{Examination and Knowledge Collection.}
\tool then examines the LLM on each exam.
When answering, the LLM is allowed to query any project knowledge it deems necessary.
The core instruction is: 
\textit{``... decide which one among the list of alternative assertion statements is equipped with a correct oracle. You must be very confident in your answer. You CANNOT guess or assume the existence or values for APIs and Fields. If you think the context lacks some important information for making the decision, please do not make a choice and output only a list of the required information ...''}
To support such queries,
\tool utilizes \texttt{CodeQL} to collect, offline, the definitions and invocations of all classes, constructors, methods, and fields in the project.
\rev{During examination, retrieval is restricted to symbols directly referenced by the focal method or the initial test (e.g., parent classes and overridden methods of the referenced symbols), ensuring semantic completeness while controlling context size.}
For example, as shown in \autoref{fig:overview}, the LLM queries the definition of \texttt{matchPath()},
because understanding its functionality is essential for selecting the correct oracle.  
\rev{Queries can be raised proactively by the LLM or enforced by \tool when the LLM selects an incorrect option, which indicates insufficient semantic understanding of $fm$.}
\rev{The examination follows an iterative loop of (i) answering, (ii) knowledge retrieval if needed, and (iii) re-evaluation, until the LLM passes the exams or a maximum number (e.g., 3) of iterations is reached.}

By examining the LLM, \tool reliably collects relevant project knowledge and strengthens the understanding of requirements and test scenarios, thereby improving subsequent test generalization (\rev{a concrete example illustrating the knowledge retrieval process is presented in Section \ref{sec:ablation-result}}).
The collected knowledge
is passed to Stage 2 for test scenario generalization.

\subsection{Test Scenario Generalization}
Given the focal method $fm$, initial test $tc$, and the project knowledge $K$ (if any) 
obtained from Stage 1, \tool generalizes test scenarios by first generating a test scenario template $so_{temp}$ and then crystallizing it into a set of test scenario instances $\{so_{ins}^m\}_{m=1}^M$ (see Section \ref{sec:test_generalization}).
To ensure comprehensive scenario coverage, the prompt for generating $so_{temp}$ is optimized using our prompt auto-tuning technique (see Section \ref{sec:approach_auto-tuning}).

\subsubsection{Test Scenario Template and Instance Generation}
\label{sec:test_generalization}
\begin{table}[t]
\setlength{\abovecaptionskip}{0pt}
\setlength{\belowcaptionskip}{0pt}
\caption{
Prompt template for generating test scenario templates. 
Green-highlighted contents are case-specific and are to be filled in accordingly. 
Gray-highlighted contents are predefined definitions and instructions. 
}
\vspace{4pt}
\label{tab:prompt-template-generation}
\centering
\footnotesize

\begin{tabularx}{\linewidth}{X}

    \toprule

    \#\underline{\textbf{Instruction}}: 
    Given the Java Focal Method, generate a Test Scenario Template for the Focal Method with reference to the provided Test Case.
    \coloredbox{SoftGray}{[Definition of a test scenario template.]}\\

    \midrule

    \#\underline{\textbf{Focal Method}}: \coloredbox{SoftGreen}{[Code of focal method]} \\

    \#\underline{\textbf{Focal Method Context}}: \coloredbox{SoftGreen}{[Skeleton of focal file]} \\

    \#\underline{\textbf{Initial Test Case}}: \coloredbox{SoftGreen}{[Developer-written test]}  \\

    \#\underline{\textbf{Project Knowledge}}: \coloredbox{SoftGreen}{[List of collected knowledge]} \\

    \midrule
    \#\underline{\textbf{Rules for Variation Point Identification}}: \coloredbox{SoftGray}{[Auto-tuned rules]} \\

    \#\underline{\textbf{Requirements}}: \\
    \ding{172} Your response must first output the analysis and thinking:
    \coloredbox{SoftGray}{[Example  analysis format]}\\

    \ding{173} Based on your analysis and thinking, if you require more Relevant Project Knowledge (e.g., constructor, method, and field) for accurately identifying variation points, output your query in the following format without any commentary:
    \coloredbox{SoftGray}{[Example query format]}\\

    \ding{174} If no more information is required, output Test Scenario Template in the following format:
    \coloredbox{SoftGray}{[Example template format]}\\
    \bottomrule
\end{tabularx}

\end{table}

Test scenario generalization is the core of \tool.
It first generates a test scenario template and then crystallizes the template into a comprehensive set of test scenario instances.

\noindent \textbf{Test Scenario Template Generation.} 
Given $fm$, $tc$, and $K$, \tool orchestrates a learnable prompt (shown in \autoref{tab:prompt-template-generation}) to guide the LLM in producing a test scenario template.
A template is defined as a general test plan that captures the developer’s intent to validate the $fm$ across diverse scenarios. 
It consists of concise steps expressed in natural language, optionally supplemented with minimal program elements.
Each step includes:
(1) \textit{Action (ACT)}: an imperative instruction for a tester.  
(2) \textit{Variation Points (VPs)}: factors that can vary, either concrete code elements or abstract objects in the high-level scenario. 
VPs can be instantiated into specific settings to produce meaningful behaviors of the focal method.   
(3) \textit{Dependency (DP)}: optional links to VPs defined in earlier steps.  

\begin{figure*}[t]
  \centering
  \includegraphics[width=1\textwidth]{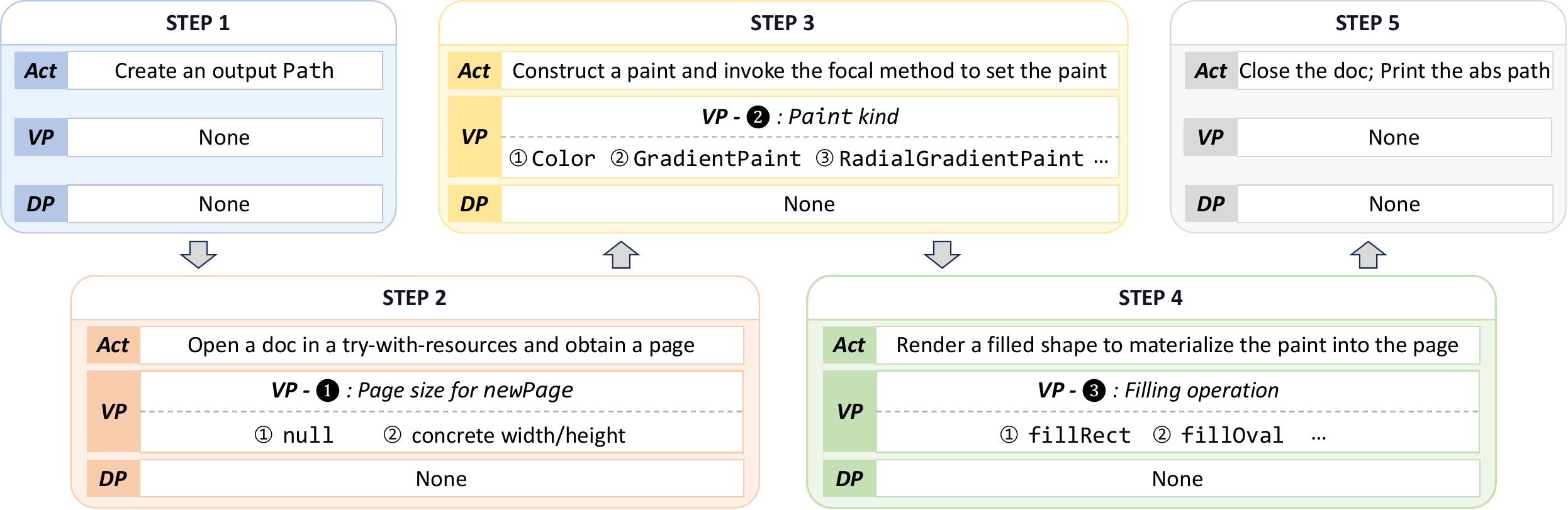}
  \caption{Generated test scenario template for \texttt{setPaint()} using \texttt{linearGradientPaint()} as the initial test.}
  \vspace{-8pt}
  
  \label{fig:template}
\end{figure*}

A high-quality template should abstract all developer-intended test scenarios and be capable of deriving them. 
For example, \autoref{fig:template} shows the template generated for the focal method \texttt{setPaint()} with \texttt{linearGradientPaint()} as the initial test.
The template consists of five steps, corresponding to the underlying pattern illustrated in \autoref{fig:motivation-1-tests}.
Among them, Steps 2–4 correctly capture the VPs and their reasonable candidate settings.  

To generate such a high-quality template, the key challenge lies in accurately identifying VPs.
To address this, the learnable prompt incorporates auto-tuned rules (``\#Rules for Variation Point Identification'' in \autoref{tab:prompt-template-generation}), 
derived through our prompt auto-tuning technique (see Section \ref{sec:auto-tuning}).
In addition, \tool encourages the LLM to proactively query additional project knowledge beyond what was retrieved in Stage 1, since some information about VPs or their feasible settings may reside in other parts of the codebase.
\rev{Such queries are resolved using the same offline CodeQL index constructed in Stage~1, and the retrieved definitions/usages are appended to the prompt before continuing template or instance generation.}

\noindent \textbf{Test Scenario Instance Generation.}
A test scenario instance is a concrete test plan for validating a specific behavior of $fm$ in a particular scenario. 
It is derived from the test scenario template by fixing all VPs and determining the associated oracles.
\chg{To instantiate a test scenario template, \tool adopts a structured instantiation strategy that separates scenario-level decision making from natural-language instance generation.
Rather than directly asking the LLM to generate full natural-language test scenario instances, \tool first asks it to output structured settings, including the selected setting of each variation point, dependencies among these settings, and the corresponding primary oracle(s).
A deterministic instantiation procedure then fills the original test scenario template with these settings to produce test scenario instances compatible with the downstream test-generation stage, while validating that all settings refer to declared variation points and valid template steps.}

Considering that $fm$ may be buggy in practice,
\tool does not assume the correctness when determining oracles. 
Instead, \tool prompts the LLM to identify and list all reasonable oracles that a developer might intend, including both the \textit{primary oracle} (deduced from the code implementation) and \textit{alternative oracles} (inferred from requirement understanding and general knowledge).  
In practice, \tool can optionally allow developers to select among alternative oracles; if no intervention occurs, the primary oracle is used.

The test scenario instances then serve as guidance for generating executable tests that comprehensively cover the developer’s intended scenarios (see Section~\ref{sec:test_generation}).

\subsubsection{Prompt Auto-Tuning}
\label{sec:approach_auto-tuning}
\begin{algorithm}[t]
\caption{Prompt Auto-Tuning for Variation Point Identification}
\label{alg:prompt_tuning}
\footnotesize
\begin{algorithmic}[1]
\STATE \textbf{Input:} (1) Black-box large language model $LLM$; (2) Dataset $D = \{(x_i,y_i)\}_{i=1}^{|D|}$, where $x_i = (fm_i, tc_i, K_i)$ bundles a focal method $fm_i$, one initial test $tc_i$, and the relevant knowledge $K_i$; $y_i$ is the ground-truth test scenario template.
\STATE \textbf{Output:} Tuned prompt $p^*$ with an optimal rule set $S_{rule}$ for VP identification.

\STATE Randomly split $D$ into training and test sets: $D_{train}$ and $D_{test}$
\STATE Randomly split $D_{train}$ into batches: $B = \{b_n\}_{n=1}^{N}$, where $b_n = \{(x^k, y^k)\}_{k=1}^{|b_n|}$
\STATE Initialize base prompt $p_0$ with empty $S_{rule}$

\FOR{epoch $e=1$ to $E$}
    \STATE $p_e \leftarrow p_{e-1}$
    \FOR{each batch $b_n \in B$}
        \STATE $T_n=\{t^k | t^k \leftarrow LLM(p_{e}, x^k)\}_{k=1}^{|b_n|}$ \COMMENT{\textit{Generate scenario templates}}
        \STATE $F_n=\{f^k|f^k\leftarrow LLM(p_{e}, y^k, t^k)\}_{k=1}^{|b_n|}$ \COMMENT{\textit{Generate feedback}}
        \STATE $p_{e}\leftarrow LLM(p_{e}, F_n)$ \COMMENT{\textit{Optimize prompt based on feedback}}
    \ENDFOR
    \STATE Record $p_e$
    
\ENDFOR
\STATE $p^* = \arg\max_{p \in \{p_1, \cdots, p_E\}}  \text{Evaluate}(p; D_{test})$  \COMMENT{\textit{Select the optimal prompt}}

\STATE \textbf{Return:} $p^*$
\end{algorithmic}
\end{algorithm}

As discussed in Section~\ref{sec:motivation} (Implicit Pattern Recognition), even with all necessary information available, accurately identifying VPs regarding test scenarios from changeable code elements remains challenging.
Many elements may appear to be plausible VPs, but only a small subset is truly meaningful.

Thus, well-designed rules are essential for accurate identification.  
However, manually crafting such rules is often time-consuming, error-prone, and lacks generalizability across projects and programming languages.  
To address this, we propose a \textit{prompt auto-tuning} technique, enabling the LLM itself to derive and optimize these rules.
Algorithm \ref{alg:prompt_tuning} outlines the procedure.
It takes as input an LLM $LLM$ and a dataset $D=\{(x_i, y_i)\}$, where $x_i = (fm_i, tc_i^j, K_i)$ bundles a focal method $fm_i$, one of its initial tests $tc_i^j$, and the relevant knowledge $K_i$.
The target $y_i$ is the ground-truth test scenario template containing accurate VPs. 
Prompt auto-tuning derives a rule set $S_{rule}$ and iteratively refines it on $D$, leveraging feedback from $LLM$ by comparing generated templates against $y_i$ to progressively improve VP identification.

\noindent \textbf{Dataset Construction.} 
We construct $D$ in a semi-automated manner. 
First, we select a set of focal methods $\{fm_i\}_{i=1}^I$ from our dataset of tests (see Section \ref{sec:dataset}).
Each selected focal method is inspected to ensure that it is equipped with sufficient developer-written tests covering comprehensive scenarios.  
Given a focal method $fm_i$ and its complete set of tests $\{tc_i^j\}_{j=1}^J$, 
we prompt the LLM to generate a test scenario template $t_i$.
Since the LLM has access to all tests, this generation task is relatively straightforward: it can infer VPs by analyzing differences among the tests.  
To validate $t_i$, we then ask the LLM to instantiate it into test scenario instances and manually compare them against $\{tc_i^j\}_{j=1}^J$. 
If missing or redundant instances are observed---indicating misidentified VPs---we provide the LLM with feedback, including the missing or redundant instances and any manually collected relevant knowledge $K_i$, and ask the LLM to refine $t_i$.  
This process is repeated until $t_i$ passes evaluation, 
at which point it is accepted as the ground-truth template.

\noindent \textbf{Tuning Process.}
The dataset $D$ is divided into a training set $D_{train}$ and a test set $D_{test}$.
$D_{train}$ is used to optimize $S_{rule}$,
while $D_{test}$ is used to select the optimal version of $S_{rule}$ across $E$ epochs.
Specifically, $D_{train}$ is partitioned into batches $B=\{b_n\}_{n=1}^{N}$.
The base prompt $p_0$ (shown in \autoref{tab:prompt-template-generation}) is initialized with an empty $S_{rule}$.
During the $e$-th epoch, the prompt is iteratively optimized across $B$:
\stepcircle{1}
For each sample $(x^k, y^k)$ in batch $b_n$, the current prompt $p_{e}$ is used to generate a template $t^k$ for $x^k$, producing a set of templates $T_n=\{t^k\}_{k=1}^{|b_n|}$.
\stepcircle{2}
$LLM$ evaluates $t^k$ against the corresponding ground-truth template $y^k$, identifying missing or redundant VPs. 
\stepcircle{3}
For $t^k$, $LLM$ generates feedback $f^k$ describing how $S_{rule}$ should be adjusted to fix the issues while preserving its generality, yielding a set of feedback $F_n=\{f^k\}_{k=1}^{|b_n|}$.
\stepcircle{4}
To avoid overfitting to particular focal methods, $LLM$ synthesizes general feedback from $F_n$ by merging complementary suggestions and resolving conflicts (i.e., retaining the most general suggestion).
This feedback is then used to refine $S_{rule}$, resulting in updated $p_{e}$.
At the end of each epoch $e$, the updated $p_e$ is recorded.

After all epochs, all prompts $\{p_e\}_{e=1}^E$ are evaluated on $D_{test}$ using precision, recall, and F1-score metrics.  
Precision measures the proportion of VPs in the generated template that also appear in the ground-truth template, while recall measures the proportion of ground-truth VPs correctly identified in the generated template.  
Algorithm \ref{alg:prompt_tuning} returns the prompt with the highest F1-score on $D_{test}$, which is then used for test scenario template generation (see Section \ref{sec:test_generalization}).

\subsection{Test Generation}
\label{sec:test_generation}
For each generated test scenario instance, \tool prompts the LLM to produce a corresponding test. 
The test is then compiled and executed, and feedback messages are collected from the compiler and runtime.
If the test fails due to compilation errors, execution errors, or assertion failures, \tool refines it based on the feedback:  
\stepcircle{1}
Extract error messages and filter out irrelevant messages that originate outside the project (e.g., third-party dependencies).
\stepcircle{2}
Use regular expressions to extract program elements and their positions from the error messages.
\stepcircle{3}
For each extracted program element, retrieve relevant project knowledge (e.g., the implementation of a method that failed during compilation or execution) via \texttt{JDTLS}~\cite{jdtls}, using the element's position for lookup.
\stepcircle{4}
\tool then constructs a prompt that integrates the focal method, the generated test, error messages, and the retrieved knowledge.
This prompt guides the LLM in refining the test to resolve the error.  
The refinement process repeats until the test passes or a maximum number of iterations is reached (e.g., three).

\section{Evaluation}
\label{sec_evaluation}
We evaluate \tool through the following research questions:

\noindent \textbf{RQ1 (Overall Performance):} How effective is \tool in generalizing tests compared to state-of-the-art baselines?

\noindent \textbf{RQ2 (Prompt Auto-Tuning Effectiveness):}
How much does prompt auto-tuning improve VP identification over standard prompting strategies?

\noindent \textbf{RQ3 (Ablation Study):} 
What are the contributions of project knowledge (i.e., retrieval) and rules (i.e., prompt auto-tuning) to \tool's overall performance?

\noindent \textbf{\rev{RQ4 (Sensitivity Analysis):}}
\rev{How robust is \tool when the initial test is of low quality?}

\noindent \textbf{\rev{RQ5} (Field Study):} 
Do the generalized tests provide practical value in real projects?

\subsection{Overall Performance (RQ1)}
\subsubsection{Dataset}
\label{sec:dataset}

\begin{table}[]
\caption{\rev{Statistics of the dataset}}
\vspace{-8pt}

\label{tab:stat_dataset}
\centering
\scriptsize
{
\color{black}
\begin{tabular}{ccccccc}
\toprule
\multirow{2}{*}{\textbf{Project}} & \multirow{2}{*}{\textbf{Commit Version}} & \multirow{2}{*}{\textbf{\begin{tabular}[c]{@{}c@{}}\# Focal\\ Methods\end{tabular}}} & \multirow{2}{*}{\textbf{\begin{tabular}[c]{@{}c@{}}\# Tests\\ (Scenarios)\end{tabular}}} & \multicolumn{3}{c}{\textbf{\# Tests per FM}} \\ \cmidrule(lr){5-7} 
                                   &                                                                                    &                                                                                      &                                                                                          & \textbf{Min}     & \textbf{Max}      & \textbf{Avg}    \\ \midrule
itext-java                         & 9c895e8410                                                                         & 177                                                                                  & 626                                                                                      & 2                & 15                & 3.5             \\
hutool                             & 3e716fcf0a                                                                         & 30                                                                                   & 73                                                                                       & 2                & 7                 & 2.4             \\
yavi                               & de9e9ab34d                                                                         & 74                                                                                   & 167                                                                                      & 2                & 8                 & 2.3             \\
lambda                             & d360ae809f                                                                         & 59                                                                                   & 144                                                                                      & 2                & 5                 & 2.4             \\
jInstagram                         & 82834b0ea2                                                                         & 29                                                                                   & 90                                                                                       & 2                & 13                & 3.1             \\
truth                              & 23171822b                                                                          & 63                                                                                   & 308                                                                                      & 2                & 15                & 4.9             \\
cron-utils                         & d31697ec4c                                                                         & 19                                                                                   & 88                                                                                       & 2                & 13                & 4.6             \\
imglib                             & 49e238162c                                                                         & 19                                                                                   & 49                                                                                       & 2                & 4                 & 2.6             \\
ofdrw                              & beeeb31c4c                                                                         & 12                                                                                   & 28                                                                                       & 2                & 4                 & 2.3             \\
RocketMQC                          & 6790ee5dfc                                                                         & 10                                                                                   & 25                                                                                       & 2                & 5                 & 2.5             \\
blade                              & ecf15b0664                                                                         & 2                                                                                    & 4                                                                                        & 2                & 2                 & 2.0             \\
spark                              & 1973e402f5                                                                         & 12                                                                                   & 35                                                                                       & 2               & 9                & 2.9             \\ \midrule
\multicolumn{2}{c}{\textbf{Total}}                                                                                           & \textbf{506}                                                                         & \textbf{1637}                                                                            & \textbf{2.0}     & \textbf{8.3}     & \textbf{3.0}    \\ \bottomrule
\end{tabular}
}
\vspace{-5pt}
\end{table}

We curate focal methods, each with developer-written tests, from 12 diverse open-source projects.
These projects span domains such as web development and image processing, each with over 100 GitHub stars and forks.

\noindent \rev{\textbf{Criteria.}
A method is selected as a focal method if it satisfies the following criteria:
\begin{itemize}[leftmargin=*]
    \item \textbf{Multiple Developer-Written Tests.}
    The method must be exercised by at least two distinct developer-written tests. 
    Our approach relies on scenario diversity to generalize developer intent.
    \item \textbf{Clear Test–Method Mapping.}
    The method must be explicitly invoked in the test body (directly or via a simple wrapper), ensuring unambiguous scenario attribution.
    \item \textbf{Meaningful Behavioral Logic.}
    We exclude trivial methods (e.g., simple getters/setters) where scenario variation is inherently limited.
    \item \textbf{Interpretable Test Scenarios.}
    Four authors inspect the associated tests to ensure they encode identifiable behavioral scenarios (e.g., nominal, boundary, or exceptional cases). 
    Ambiguous cases are resolved through discussion.
\end{itemize}
}

\noindent \rev{\textbf{Statistics.}
\autoref{tab:stat_dataset} presents the commit versions of the projects in which the focal methods are selected, the number of focal methods, the total number of existing tests for these focal methods, and the minimal, maximal, and average number of existing tests per focal method.
}
In total, the dataset contains 506 focal methods and 1,637 tests.
Tests range from 13 to 140 lines of code (LOC; average 37), while focal methods range from 3 to 90 LOC (average 11).
\rev{For RQ1, we evaluate \tool on seven projects (see \autoref{tab:rq1_overall}). 
The remaining five projects are exclusively used for prompt auto-tuning in RQ2.}

\subsubsection{Baselines}
We use gpt-o4-mini (\texttt{o4-mini-2025-04-16}) as the default LLM for \tool.
To assess practicality with an open-source model, we also report \tool powered by DeepSeek-V3.1 (denoted ``Ours (ds)''). 
For comparison, we evaluate three representative baselines:
\textbf{\ding{172} o4-mini}, a vanilla baseline, 
\textbf{\ding{173} ChatTester~\cite{chattester}}, a state-of-the-art LLM-based test generator, and 
\textbf{\ding{174} EvoSuite~\cite{fraser2011evosuite}}, a state-of-the-art search-based test generator.

We use o4-mini as a vanilla baseline, as it is a straightforward tool for test generalization.
ChatTester is chosen for its outstanding performance among LLM-based test generators.
For fairness, we upgrade its backend from GPT-3.5 to o4-mini.
To adapt ChatTester to the test generalization task, we provide it with an initial test and modify its prompt from ``generate one test case'' to ``generate more test cases with reference to the initial test case'', leaving other settings unchanged.

EvoSuite is chosen as a representative search-based technique.
As a coverage-driven generator, it naturally produces multiple tests for a given focal method, making it a suitable non-LLM baseline.

\subsubsection{Metrics}
We evaluate \tool and the baselines using \textit{scenario coverage}, which measures how well the generated tests cover the test scenarios targeted by developer-written tests.  
For a focal method $fm^n \in S_{fm}$ with generated tests $S_{gen}^n$ and ground-truth tests $S_{gt}^n$, scenario coverage is defined as:
{
\small
\begin{align}
    Cov_{so}^n = \frac{1}{|S_{gt}^n|}\sum_{tc_{gt} \in S_{gt}^n} \mathrm{Coverage}(tc_{gt}, S_{gen}^n).
    \label{eq:cov}
\end{align}
}
We adopt two complementary definitions of $\mathrm{Coverage}(\cdot,\cdot)$:  
\begin{itemize}[leftmargin=*]  
  \item \textbf{Mutation-based Scenario Coverage.} 
  The intuition is that a scenario targeted by a test can be characterized by the set of potential bugs it can expose.
  We utilize Pitest~\cite{pitest} to mutate the class containing $fm^n$, yielding a mutant set.
  For a test $tc$, let $S_{\mu}(tc)$ denote the mutants killed by $tc$.
  For a ground-truth test $tc_{gt}$ and a generated test $tc_{gen}$, the pairwise coverage score is
  {\small
  \begin{align}
      s(tc_{gt}, tc_{gen}) = \frac{|S_{\mu}(tc_{gt}) \cap S_{\mu}(tc_{gen})|}{|S_{\mu}(tc_{gt})|}.
  \end{align}
  }
  To ensure \textit{order-independent} and prevent a single generated test from matching multiple ground-truth tests, we compute a maximum-weight bipartite matching between $S_{gt}^n$ and $S_{gen}^n$ with edge weights $s(\cdot,\cdot)$, and the matched scores (unmatched $tc_{gt}$ contribute $0$) are then used in \autoref{eq:cov}.
  \item \textbf{LLM-Assessed Scenario Coverage.}
  Mutation-based scenario coverage can be overly strict and miss semantically valid matches.
  For example, testing a default behavior of a focal method either by omitting a setting or by explicitly resetting it may kill different mutants while targeting the same scenario.
  Moreover, a developer-written test may encode multiple scenarios, whereas generated tests target a single scenario.
  Therefore, LLM-assessed scenario coverage serves as a complementary metric.
  We prompt the LLM \rev{(same as that used for test generalization)} to judge whether the test scenario(s) exercised by $tc_{gt}$ can be fulfilled by any tests in $S_{gen}^n$.  
  If so, $\mathrm{Coverage}(tc_{gt}, S_{gen}^n) = 1$; otherwise it is $0$.
  To avoid double counting, the tests decided to match $tc_{gt}$ are removed from $S_{gen}^n$ before evaluating the next ground-truth test.
  
\end{itemize}

\subsubsection{Results}
\begin{table}[]
\caption{\rev{Comparison of \tool and baselines in test scenario coverage (in \%).}}
\vspace{-8pt}

\label{tab:rq1_overall}
\centering
\resizebox{\textwidth}{!}{
\color{black}
\begin{tabular}{ccrrrrrrrrrr}
\toprule
                                            &                                         & \multicolumn{5}{c}{\textbf{Mutation-based Scenario Coverage}}                                                                                                                                       & \multicolumn{5}{c}{\textbf{LLM-Assessed Scenario Coverage}}                                                                                                                                    \\ \cmidrule(lr){3-7} \cmidrule(lr){8-12}
\multirow{-2}{*}{\textbf{Projects}}         & \multirow{-2}{*}{\textbf{\# Scenarios}} & \multicolumn{1}{c}{\textbf{EvoSuite}} & \multicolumn{1}{c}{\textbf{Vanilla}} & \multicolumn{1}{c}{\textbf{ChatTester}} & \multicolumn{1}{c}{\textbf{Ours (ds)}} & \multicolumn{1}{c}{\textbf{Ours}} & \multicolumn{1}{c}{\textbf{EvoSuite}} & \multicolumn{1}{c}{\textbf{Vanilla}} & \multicolumn{1}{c}{\textbf{ChatTester}} & \multicolumn{1}{c}{\textbf{Ours (ds)}} & \multicolumn{1}{c}{\textbf{Ours}} \\ \midrule
\textbf{itext-java} & 627                                     & 8.53                                  & 31.91                                & 30.88                                   & \cellcolor[HTML]{EFEFEF}70.57                                  & 65.52                              & 19.28                                 & 33.86                                & 32.36                                   & 56.14                                  & \cellcolor[HTML]{EFEFEF}57.95                             \\
\textbf{hutool}     & 73                                      & 4.39                                  & 65.33                                & 51.19                                   & 81.79                                  & \cellcolor[HTML]{EFEFEF}84.07                              & 4.88                                  & 37.67                                & 44.75                                   & 69.97                                  & \cellcolor[HTML]{EFEFEF}70.21                             \\
\textbf{lambda}     & 144                                     & \textit{n/a}                                     & 65.86                                & 76.89                        & \cellcolor[HTML]{EFEFEF}96.05          & 95.49                              & \textit{n/a}                                     & 57.99                                & 65.74                                   & \cellcolor[HTML]{EFEFEF}78.76                                  & 77.89                             \\
\textbf{cron-utils}                         & 88                                      & 16.32                                 & 26.81                                & 29.38                                   & \cellcolor[HTML]{EFEFEF}72.46                                  & 64.75                              & 15.79                                 & 23.81                                & 35.83                                   & \cellcolor[HTML]{EFEFEF}58.16                                  & 57.91                             \\
\textbf{ofdrw}                              & 28                                      & \textit{n/a}                                     & 12.21                                & 14.89                                   & \cellcolor[HTML]{EFEFEF}75.28                                  & 42.14                              & \textit{n/a}                                     & 26.79                                & 23.57                                   & 47.62                                  & \cellcolor[HTML]{EFEFEF}64.88                             \\
\textbf{RocketMQC}                          & 25                                      & 25.99                                 & 58.86                                & 64.86                                   & 82.37                                  & \cellcolor[HTML]{EFEFEF}89.75                              & 29.00                                 & 54.00                                & 75.00                                   & 81.00                                  & \cellcolor[HTML]{EFEFEF}85.00                             \\
\textbf{blade}                              & 4                                       & 42.67                                 & 17.67                                & 51.00                                   & 84.33                                  & \cellcolor[HTML]{EFEFEF}99.00                              & 0.00                                  & 50.00                                & 75.00                                   & \cellcolor[HTML]{EFEFEF}100.00                                 & \cellcolor[HTML]{EFEFEF}100.00                            \\ \midrule
\textbf{Average}           & \textbf{989}                   & \textbf{19.58}               & \textbf{39.81}              & \textbf{45.58}                 & \cellcolor[HTML]{EFEFEF}\textbf{80.41} & \textbf{77.25}           & \textbf{13.79}               & \textbf{40.59}              & \textbf{50.32}                 & \textbf{70.24}                & \cellcolor[HTML]{EFEFEF}\textbf{73.41} \\ \bottomrule
\end{tabular}
}
\vspace{-12pt}
\end{table}

\autoref{tab:rq1_overall} reports the results of \tool and baselines in terms of both mutation-based and LLM-assessed scenario coverage across \rev{seven projects and 989} test scenarios.
On average, a ground-truth test scenario involves \rev{82.1} mutants.
The column ``Ours (ds)'' shows results when \tool is powered by DeepSeek-V3.1, while the column ``Ours'' shows \tool's performance when powered by gpt-o4-mini.
For projects \texttt{ofdrw} and \texttt{lambda},
the results of EvoSuite are marked as ``\textit{n/a}'' because EvoSuite fails to generate tests for most focal methods in these projects, which involve extensive use of generic Java types.   
The last row ``Average'' presents the average performance across all projects, with the projects where ``\textit{n/a}'' values occur excluded for EvoSuite.

\noindent \textbf{Baseline Comparison \rev{in Scenario Coverage.}}
Overall, EvoSuite, despite its efficiency in achieving high code coverage, performs far worse than both LLM-based baselines and \tool.
For example, its mutation-based scenario coverage is lower by \rev{20.23\%} than Vanilla and by \rev{57.67\%} than \tool.
These results are expected:
Code coverage-driven techniques rely entirely on control-flow branches in the focal method, but scenarios do not necessarily correspond to branches, and many focal methods contain no branches.
For \texttt{setPaint} (introduced in Section~\ref{sec:motivation}), once a single test is generated, such approaches terminate, leaving many important scenarios uncovered.
By contrast, LLM-based baselines (i.e., Vanilla and ChatTester) achieve much better performance.
Their advantage comes from the ability of LLMs to infer or \textit{``imagine''} plausible scenarios beyond code structural exploration.  
For example, as shown in \autoref{fig:motivation-1-chattester}, ChatTester can imagine a test scenario where the paint is set to a basic color using a \texttt{Color} object.

However, imagination requires both constraint and stimulation; without them, it can lead to \textit{hallucination} and \textit{exhaustion}.
Hallucination occurs when a tool treats unintended factors as VPs.
For example, in project \texttt{spark}~\cite{spark}, ChatTester treats the parameters \textit{StaticFilesConfiguration} and \texttt{ExceptionMapper} of focal method 
\texttt{EmbeddedJettyFactory.create()}
as VPs,
generating tests that vary them (e.g., setting both to \texttt{null}).
Yet the initial test uses \texttt{mock} objects for these parameters, clearly indicating that developers do not intend their effects to be part of the test scenarios.  

Exhaustion occurs when a tool fails to derive a comprehensive set of reasonable scenarios, 
even after correctly identifying VPs.
As shown in \autoref{fig:motivation-1-chattester}, ChatTester generated only one reasonable scenario (\texttt{testSetPaintWithBasicColor()}) regarding a VP (i.e., the parameter of \texttt{setPaint()}), while missing others intended by the developer, such as \texttt{setPaintRadialGradientPaint()}.  

These issues stem from the lack of explicit rules and project-specific knowledge to constrain VP identification and to stimulate exploration of reasonable settings.
\tool addresses both: it combines rule-based guidance with retrieved project knowledge and therefore consistently outperforms LLM-based baselines.
Overall, \tool achieves \rev{77.25\%} mutation-based scenario coverage and \rev{73.41\%} LLM-assessed scenario coverage, 
improving over Vanilla by \rev{37.44\%} and \rev{32.82\%}, and over ChatTester by \rev{31.66\%} and \rev{23.08\%}, respectively.
A mutation-based scenario coverage \rev{close to} 80\% indicates that the generalized scenarios are sufficiently close to the ground truth, given the large number of mutants per ground-truth scenario on average (\rev{82.1}).

\noindent \textbf{Mutation-based vs. LLM-Assessed Scenario Coverage.}
Comparing mutation-based and LLM-assessed scenario coverage, we observe moderate differences between them. 
For example, mutation-based scenario coverage is often slightly higher than LLM-assessed scenario coverage (e.g., on project \rev{\texttt{RocketMQC}}).
This is because mutation-based coverage provides a continuous score in $[0,1]$ that reflects partial overlap in killed mutants,
whereas LLM-assessed coverage is binary (1 if any generated tests fulfill the scenario of the ground-truth test, 0 otherwise). 
Mutation-based coverage can capture partial alignment between two tests, while LLM-assessed scenario coverage cannot.

\begin{table}[]
\caption{\rev{Branch and line coverage achieved by \tool (powered by gpt-o4-mini) and baselines.}}
\vspace{-8pt}

\label{tab:cov_project}
\centering
\footnotesize
\color{black}
\begin{tabular}{ccccccccc}
\toprule
                                   & \multicolumn{4}{c}{\textbf{Branch Coverage}}                                                                                & \multicolumn{4}{c}{\textbf{Line Coverage}}                                                                                  \\ \cmidrule(lr){2-5} \cmidrule(lr){6-9} 
\multirow{-2}{*}{\textbf{Project}} & \textbf{EvoSuite}             & \textbf{Vanilla}             & \textbf{ChatTester} & \textbf{Ours}                          & \textbf{EvoSuite}             & \textbf{Vanilla}             & \textbf{ChatTester} & \textbf{Ours}                          \\ \midrule
itext-java                         & \cellcolor[HTML]{EFEFEF}28.27  & 8.23                          & 14.19                          & 21.02                                  & \cellcolor[HTML]{EFEFEF}35.63    & 14.05            & 19.65                           & 29.22                                  \\
hutool                             & 3.33                           & 3.97                          & \cellcolor[HTML]{EFEFEF}6.89   & 5.43                                   & 4.98                             & 3.70             & \cellcolor[HTML]{EFEFEF}8.50    & 6.23                                   \\
lambda                             & -                              & 38.50                         & 38.18                          & \cellcolor[HTML]{EFEFEF}40.10          & -                                & 28.21            & 27.52                           & \cellcolor[HTML]{EFEFEF}29.22          \\
cron-utils                         & 10.87                          & 31.89                         & 30.93                          & \cellcolor[HTML]{EFEFEF}36.66          & 22.34                            & 43.96            & 43.50                           & \cellcolor[HTML]{EFEFEF}47.17          \\
ofdrw                              & -                              & 1.53                          & 3.02                           & \cellcolor[HTML]{EFEFEF}13.71          & -                                & 4.35             & 6.73                            & \cellcolor[HTML]{EFEFEF}21.68          \\
RocketMQC                          & 1.24                           & 1.22                          & 1.26                           & \cellcolor[HTML]{EFEFEF}1.85           & 5.08                             & 0.73             & 4.87                            & \cellcolor[HTML]{EFEFEF}5.93           \\
blade                              & 5.80                           & 6.40                          & 6.40                           & \cellcolor[HTML]{EFEFEF}6.70           & 10.14                            & 11.12            & 11.12                           & \cellcolor[HTML]{EFEFEF}11.30          \\ \midrule
\textbf{Average}                   & \textbf{9.90}                  & \textbf{13.11}                & \textbf{14.41}                & \cellcolor[HTML]{EFEFEF}\textbf{17.92} & \textbf{15.63}                   & \textbf{15.16}   & \textbf{17.41}                  & \cellcolor[HTML]{EFEFEF}\textbf{21.54} \\ \bottomrule
\end{tabular}
\vspace{-5pt}
\end{table}

\begin{wraptable}[11]{r}{0.55\textwidth}
\caption{\rev{Mutation score on the whole project achieved by \tool (powered by gpt-o4-mini) and baselines.}}
\vspace{-6pt}

\label{tab:mut_project}
\centering
\scriptsize
{
\color{black}
\begin{tabular}{ccccc}
\toprule
\textbf{Project}    & \textbf{EvoSuite} & \textbf{Vanilla}                       & \textbf{ChatTester} & \textbf{Ours}                          \\ \midrule
itext-java & 11.71                                     & 28.41              & 32.86                            & \cellcolor[HTML]{EFEFEF}43.44 \\
hutool     & 5.59                                      & 12.39              & \cellcolor[HTML]{EFEFEF}17.13    & 15.53                         \\
lambda     & -                                         & 56.85              & 60.67                            & \cellcolor[HTML]{EFEFEF}62.25 \\
cron-utils & 23.64                                     & 59.74              & 71.86    & \cellcolor[HTML]{EFEFEF}71.88 \\
ofdrw      & -                                         & 11.73              & 12.29                            & \cellcolor[HTML]{EFEFEF}26.82 \\
RocketMQC  & 16.37                                     & 27.19              & 45.03                            & \cellcolor[HTML]{EFEFEF}51.46 \\
blade      & 31.88                                     & 48.47              & 50.22                            & \cellcolor[HTML]{EFEFEF}87.77 \\ \midrule
\textbf{Average}    & \textbf{17.84}                   & \textbf{34.97}  & \textbf{41.44}                      & \cellcolor[HTML]{EFEFEF}\textbf{51.31} \\ \bottomrule
\end{tabular}
}
\end{wraptable}

On the other hand, LLM-assessed coverage is more flexible when a ground-truth test encodes multiple scenarios.  
For example, in project \texttt{ofdrw}~\cite{ofdrw}, the focal method \texttt{setClip} has a test \texttt{clip()} that exercises three distinct clipping scenarios:  
(1) a fill entirely inside the clip region,  
(2) a fill partially overlapping the region, and  
(3) a fill completely outside the region.  
\tool correctly generalized these into three separate tests.  
In this case, LLM-assessed coverage recognizes that the three generated tests collectively fulfill the test scenario targeted by the ground-truth test (i.e., $\mathrm{Coverage}(tc_{gt}, S_{gen}^n) = 1$ in \autoref{eq:cov}), while mutation-based coverage assigns only partial overlap ($\mathrm{Coverage}(tc_{gt}, S_{gen}^n) = 0.33$).

Finally, the column ``Ours (ds)'' reports the performance of \tool powered by DeepSeek-V3.1.
As we can see, \tool still consistently outperforms all baselines across all metrics, demonstrating its robustness and practicality for deployment in the industry.

\noindent \textbf{\rev{Baseline Comparison in Code Coverage and Mutation Score.}}
\rev{We also measure traditional code coverage and mutation score for a comprehensive comparison.
\autoref{tab:cov_project} and \autoref{tab:mut_project} compare the total code coverage and mutation score achieved across the whole project with all test cases generated by each approach.
\tool achieves the highest average branch coverage, line coverage, and mutation score, with gains of 3.51\%, 4.13\%, and 9.87\% over the best baseline, ChatTester.
We observe that certain individual tests generated by baselines achieve higher code coverage or mutation score than \tool.
We manually inspected these cases and found that the baseline tests often exercise additional classes or methods that are unnecessary for testing the meaningful scenarios, thereby inflating project-level coverage. 
}

\noindent \rev{\textbf{Improvement Over Existing Tests.}
We analyze how generated tests improve existing test suites. Specifically, we measure the incremental mutation and coverage gains obtained by augmenting the original tests ($T_{exist}$) with generated tests. For each project, we compare $T_{exist}$ with $T_{exist} + T_{ours}$ and compute the additional mutants killed and additional branch and line coverage achieved.}

\rev{Across all projects, augmenting the existing tests with \tool yields an average increase of 7.07\% (57.93\% to 65.00\%) in mutation score, 2.46\% (17.47\% to 19.93\%) in branch coverage, and 2.05\% (21.45\% to 23.50\%) in line coverage. 
Although the improvements are moderate --- since the focal methods in our benchmark are already supported by well-developed test suites and \tool operates from a single initial test --- these results indicate that \tool can still introduce fault-detection capability and structural coverage beyond the original developer-written tests. 
This suggests that scenario-driven generalization complements existing suites by systematically exploring intention-aligned variations rather than merely duplicating already-tested behaviors.}

\noindent \rev{\textbf{Token and Time Consumption.}
We measure the token usage and wall-clock time required to generate one test. 
Across all generations, \tool consumes on average 12K tokens per test.
The average end-to-end processing time per test is 2 minutes.
Future work could reduce cost by partially replacing LLM calls in Stage 1 with static or mutation-based program analysis.}

\subsection{Prompt Auto-Tuning Effectiveness (RQ2)}
\label{sec:auto-tuning}

\subsubsection{Baselines}
Well-designed rules are crucial for generating high-quality test scenario templates with accurate variation points.
\tool derives these rules through a prompt auto-tuning technique.
We evaluate its effectiveness by comparing it against three standard prompting strategies:
\begin{itemize}[leftmargin=*]  
  \item \textbf{Zero-Shot Prompting: }
  The LLM generates templates using a prompt with only the task description, without rules or examples for guidance.
  \item \textbf{Few-Shot Prompting:}
  The LLM is provided with one to three in-context examples, each consisting of a focal method, an initial test, and the corresponding ground-truth template.
  \item \textbf{Hand-Crafted Prompt:}
  A prompt with manually designed rules, created by annotators based on their experience when constructing the dataset for prompt auto-tuning.  
\end{itemize}

\subsubsection{Dataset}
We construct a dataset of 50 samples (see Section \ref{sec:auto-tuning}), each consisting of a focal method, its ground-truth tests, and the ground-truth template.
These focal methods are selected from \texttt{spark}, \texttt{imglib}, \texttt{truth}, \texttt{jInstagram}, and \texttt{yavi} projects, 
which span diverse scales and domains and are not used in our field study.
Each focal method is manually inspected to ensure that it is equipped with sufficient test cases covering comprehensive scenarios.
On average, a focal method has 5.6 tests (ranging from 2 to 21), and a template has 3.2 variation points (ranging from 1 to 6).

\subsubsection{Setup}
\begin{wraptable}[8]{r}{0.5\textwidth}
\caption{Comparison of prompt auto-tuning and baseline strategies for variation point identification (in \%).}
\vspace{-8pt}

\label{tab:auto-tuning}
\centering
\scriptsize
\begin{tabular}{cccc}
\toprule
\textbf{Method}              & \textbf{Precision} & \textbf{Recall} & \textbf{F1}\\ \midrule
Zero-Shot Prompting  &  49.55                &  75.49      &     59.83       \\
Few-Shot Prompting  &  51.64                &  70.24      &   59.52         \\
Hand-Crafted Prompt &    68.47              &   77.29     &   72.61         \\
Auto-Tuned Prompt  &   \cellcolor[HTML]{EFEFEF}\rev{81.21}               &   \cellcolor[HTML]{EFEFEF}\rev{86.39}     &    \cellcolor[HTML]{EFEFEF}\rev{83.72}       \\ \bottomrule
\end{tabular}
\end{wraptable}

We evaluate prompt auto-tuning and baselines using precision, recall, and F1-score for variation point identification.
For zero-shot prompting and the hand-crafted prompt, all 50 samples are used for evaluation.
\rev{For prompt auto-tuning, we apply the Leave-One-Project-Out strategy: for each fold, training on four projects and testing on the remaining one project.
We report the average evaluation results across the five folds. }
Regarding the tuning hyperparameters, we set the number of epochs to 3 and the batch size to 5.
For few-shot prompting, we randomly select 2 samples as in-context references,
and the rest of the samples are used for evaluation.
To account for variability due to reference selection, we repeat the experiment three times with different references and report the average performance across runs.

\subsubsection{Results}
\autoref{tab:auto-tuning} reports the results for each prompting strategy.
Zero-shot and few-shot prompting yield the worst performance across all metrics.
Notably, few-shot prompting performs even worse than zero-shot prompting (see Recall and F1-score metrics),
because a small number of references provides only partial guidance or causes overfitting, 
while adding more references can degrade performance due to context length constraints (the 3-reference setting performs even worse).
Hand-crafted prompting outperforms zero- and few-shot prompting but remains limited, as designing comprehensive and non-conflicting rules for complex scenarios is difficult for humans.  
In contrast, prompt auto-tuning can synthesize and reconcile rules from diverse samples into a unified, conflict-resolved rule set, achieving the best results across all metrics with improvements of \rev{31.66\%, 10.90\%, and 23.89\%} in precision, recall, and F1-score over zero-shot prompting.
These findings demonstrate the effectiveness of the proposed prompt auto-tuning technique.

\subsection{Ablation Study (RQ3)}
\label{sec:ablation}
\subsubsection{Setup}
To assess the contribution of project knowledge (i.e., retrieval), we remove all retrieved knowledge from the prompts used in Stage 2 and Stage 3.
We then re-run the full pipeline (\textit{w/o knowledge}).
As for the contribution of rules (i.e., prompt auto-tuning), we remove them from the prompt used in Stage 2 (i.e., template generation) and re-run the pipeline (\textit{w/o rules}).

\begin{table}[]
\caption{Ablation study results for \tool (in \%).}
\vspace{-8pt}
\label{tab:rq2_ablation}
\centering
\scriptsize
\begin{tabular}{ccccccc}
\toprule
                                            & \multicolumn{3}{c}{\textbf{Mutation-based Scenario Coverage}} & \multicolumn{3}{c}{\textbf{LLM-Assessed Scenario Coverage}}  \\ \cmidrule(lr){2-4} \cmidrule(lr){5-7}
\multirow{-2}{*}{\textbf{Projects}}         & \textbf{w/o knowledge}  & \textbf{w/o rules} & \textbf{Ours}  & \textbf{w/o knowledge} & \textbf{w/o rules} & \textbf{Ours}  \\ \midrule
\textbf{itext-java} & 64.78                   & 57.17              & 65.52          & 54.32                  & 53.21              & 57.95          \\
\textbf{hutool}     & 81.20                   & 77.05              & 84.07          & 67.43                  & 63.51              & 70.21          \\
\textbf{yavi}       & 91.63                   & 83.09              & 96.92          & 77.84                  & 70.34              & 81.69          \\
\textbf{lambda}     & 93.97                   & 95.83              & 95.49          & 71.06                  & 78.18              & 77.89          \\
\textbf{jInstagram} & 74.54                   & 80.12              & 77.28          & 63.81                  & 64.70              & 64.48          \\
\textbf{truth}      & 62.27                   & 41.53              & 66.25          & 59.94                  & 43.69              & 63.24          \\
\textbf{cron-utils}                         & 59.42                   & 55.73              & 64.75          & 54.89                  & 43.65              & 57.91          \\
\textbf{imglib}                             & 94.76                   & 90.83              & 94.76          & 93.20                  & 87.41              & 93.20          \\
\textbf{ofdrw}                              & 32.76                   & 33.31              & 42.14          & 39.21                  & 41.94              & 64.88          \\
\textbf{RocketMQC}                          & 84.37                   & 88.24              & 89.75          & 74.00                  & 79.00              & 85.00          \\
\textbf{blade}                              & 83.00                   & 98.67              & 99.00          & 100.00                 & 100.00             & 100.00         \\
\textbf{spark}                              & 81.48                   & 77.15              & 85.53          & 67.50                  & 70.12              & 72.19          \\ \midrule
\textbf{Average}                            & \textbf{75.35}          & \textbf{73.23}     & \textbf{80.12} & \textbf{68.60}                  & \textbf{66.31}     & \textbf{74.05} \\ \bottomrule
\end{tabular}
\vspace{-5pt}
\end{table}

\subsubsection{Results}
\label{sec:ablation-result}
\autoref{tab:rq2_ablation} reports the ablation results.
The columns ``\textit{w/o knowledge}'' and ``\textit{w/o rules}'' show the performance when project knowledge and auto-tuned rules are removed, respectively.

\noindent \textbf{Contribution of Project Knowledge.}
As shown in the ``\textit{w/o knowledge}'' column, compared to \tool, scenario coverage drops by 4.77\% and 5.45\% on average in mutation-based and LLM-assessed metrics, respectively.  
This highlights the importance of project knowledge collected across the three stages.  

\paragraph{Knowledge from Stage 1}
In project \texttt{spark}~\cite{spark}, the focal method \texttt{matches(Http\-Method http\-Method, String path)} processes the parameter \texttt{httpMethod} directly but delegates \texttt{path} handling to \texttt{matchPath(String path)}, which distinguishes four patterns (i.e., \texttt{/path}, \texttt{/path/}, \texttt{/path/other}, \texttt{/path/*}).  
Without the implementation of \texttt{match\-Path} in context, the LLM cannot fully infer the functional requirement behind the focal method, and thus generalizes tests only over \texttt{httpMethod}, missing four test scenarios related to \texttt{path}.  
For this example, \tool proactively queries the implementation of \texttt{match\-Path} in Stage 1, as it recognizes that it is crucial for determining the oracle and answering the exam correctly.  

\paragraph{Knowledge from Stage 2}
In Stage 2, project knowledge again proves essential.  
For example, for \texttt{set\-Paint(Paint paint)} from \texttt{ofdrw}~\cite{ofdrw}, the initial test \texttt{linear\-Gradient\-Paint()} contains no assertions, so Stage 1 is skipped.  
\tool prompts the LLM to proactively query project knowledge.  
Since the parameter \texttt{p} is of type \texttt{Linear\-Gradient\-Paint}, a subtype of \texttt{Paint}, the LLM queries the family of related types and thus is aware of \texttt{Color}, \texttt{Gradient\-Paint}, \texttt{Radial\-Gradient\-Paint}, and \texttt{Texture\-Paint}.  
This knowledge guides the LLM to infer comprehensive test scenarios.  

As for the project knowledge collected based on error messages, it is useful for correcting compilation and execution errors during test generation.
In Stage 3, project knowledge collected from error messages is useful for resolving compilation and execution errors.  
Since this effectiveness is well established in prior work~\cite{chattester}, we omit the discussion here.

\noindent \textbf{Contribution of Auto-Tuned Rules.}
Comparing the column ``\textit{w/o rules}'' with the column ``\tool'',
we observe that scenario coverage decreases by 6.89\% and 7.74\% on average in mutation-based and LLM-assessed metrics, respectively.  
This confirms the effectiveness of auto-tuned rules in test scenario generalization.  

Without the rules, the LLM tends to indiscriminately treat changeable details as variation points, producing inferior test scenario templates where the valuable variation points are overwhelmed by the meaningless ones.
On the one hand, redundant test scenarios derived from these spurious variation points generate noise, forcing developers to spend extra effort distinguishing useful tests.  
On the other hand, these redundant test scenarios hinder the derivation of valuable test scenarios, given the limited context length and the constrained attention capacity of the LLM.

For example, for \texttt{EmbeddedJettyFactory.create()}~\cite{spark}, the generated template incorrectly treats \texttt{StaticFilesConfiguration} and \texttt{ExceptionMapper} as variation points, even though the initial test mocks these parameters—explicitly signaling that their behavior is irrelevant to the intended scenarios.  
Similarly, trivial parameters such as connection timeouts, whose values are determined by earlier variation points, are wrongly considered independent variation points.  
As a result, the template includes eight meaningless variation points and only two actual ones, resulting in four redundant scenarios and missing one of the three intended scenarios.

\subsection{\rev{Sensitivity Analysis (RQ4)}}
\label{sec:sensitivity}

\subsubsection{\rev{Setup}}

\rev{To evaluate the robustness of \tool under degraded initial test quality, we progressively remove semantic information from the original initial test while preserving executability and the focal method invocation. }
\rev{We define three quality levels:}
\begin{itemize}[leftmargin=*]
\item \rev{\textbf{L0 (Original test).} The developer-written test is used as-is. This serves as the baseline.}

\item \rev{\textbf{L1 (Oracle-removed).} All oracle statements are removed via deterministic AST-based transformation. 
We delete assertion and verification calls (e.g., \texttt{assertEquals}, \texttt{assertTrue}, \texttt{assertThat}, \texttt{verify}), while preserving the test setup and focal method invocation. }

\item \rev{\textbf{L2 (Smoke test).} Starting from the L1 result, we use an LLM to further simplify the test into a minimal smoke test. The LLM is instructed to: (1) rename the test method to a generic name (e.g., \texttt{test\_focal\_method}), 
(2) remove scenario-specific setup code and retain only the minimal statements required to instantiate necessary objects and invoke the focal method once, and 
(3) maintain executability. 
The LLM receives only the L1 test and the focal method signature as input (not the original L0 test), to avoid reintroducing removed oracle information. }
\end{itemize}

\subsubsection{\rev{Results}}
\begin{wraptable}[12]{r}{0.5\textwidth}
\caption{\rev{Performance comparison of \tool across three levels of initial test quality (in \%).}}
\vspace{-8pt}

\label{tab:quality}
\centering
\scriptsize
{
\color{black}
\begin{tabular}{ccccccc}
\toprule
\multirow{2}{*}{\textbf{Project}} & \multicolumn{3}{c}{\textbf{Mutation-Based}} & \multicolumn{3}{c}{\textbf{LLM-Assessed}}     \\ \cmidrule(lr){2-4} \cmidrule(lr){5-7} 
                                  & \textbf{L0}    & \textbf{L1}   & \textbf{L2}   & \textbf{L0}   & \textbf{L1}   & \textbf{L2}   \\ \midrule
itext-java                        & 65.5           & 55.9          & 55.9          & 58.0          & 53.5          & 52.2          \\
hutool                            & 84.1           & 75.3          & 67.9          & 70.2          & 52.7          & 43.3          \\
lambda                            & 95.5           & 89.9          & 90.0          & 77.9          & 70.6          & 68.5          \\
cron-utils                        & 64.8           & 56.0          & 50.4          & 57.9          & 52.1          & 48.8          \\
ofdrw                             & 42.1           & 23.6          & 9.1           & 64.9          & 63.1          & 60.1          \\
RocketMQC                         & 89.8           & 80.2          & 76.9          & 85.0          & 80.0          & 73.0          \\
blade                             & 99.0           & 74.7          & 76.3          & 100           & 100           & 100           \\ \midrule
\textbf{Average}                  & \textbf{77.2}  & \textbf{65.1} & \textbf{60.9} & \textbf{73.4} & \textbf{67.4} & \textbf{63.7} \\ \bottomrule
\end{tabular}
}
\end{wraptable}

\rev{\autoref{tab:quality} reports the mutation-based and LLM-assessed scenario coverage across the seven projects evaluated in RQ1 for all three quality levels. 
On average, L0 achieves 77.2\% / 73.4\% scenario coverage.
L1 (oracle removed) achieves 65.1\% / 67.4\%, corresponding to a decrease of 12.1\% / 6.0\%. 
L2 (smoke test) achieves 60.9\% / 63.7\%, corresponding to a decrease of 16.3\% / 9.7\% from L0.
The degradation from L0 to L1 is moderate, indicating that \tool does not rely solely on explicit oracle statements; substantial scenario signal remains embedded in the test setup and invocation structure. 
While the larger drop from L0 to L2 confirms that lower-quality inputs affect performance, it also demonstrates graceful degradation. 
Even when provided with a minimal smoke test, \tool maintains over 60\% scenario coverage, indicating the pipeline remains highly effective without requiring rich initial tests.
}

\subsection{Field Study \rev{(RQ5)}}
\label{sec:field_study}

While the previous research questions evaluate \tool on datasets of developer-written tests, its practical effectiveness in real projects remains to be assessed.
In particular, some generalized tests extend beyond the ground-truth tests, raising the question of whether they are redundant or capture meaningful scenarios that developers intended but overlooked.
To investigate, we conduct a field study by submitting generalized tests as pull requests to active GitHub projects and recording the number of accepted, rejected, and pending submissions.  

\begin{wraptable}{r}{0.47\textwidth}
\caption{Results from a field study of five pull requests, comprising 27 tests for 10 focal methods.}
\vspace{-5pt}
\label{tab:field_study}
\centering
\scriptsize
\begin{tabular}{cccc}
\toprule
\textbf{Project} & \textbf{\# Accepted} & \textbf{\# Pending} & \textbf{\# Rejected} \\ \midrule
hutool           & 15                   & 0                   & 0                    \\
cron-utils       & 0                    & 9                   & 0                    \\
ofdrw            & 1                    & 1                   & 0                    \\
blade            & 0                    & 1                   & 0                    \\ \midrule
\textbf{Total}   & \textbf{16}          & \textbf{11}         & \textbf{0}           \\ \bottomrule
\end{tabular}
\end{wraptable}

Specifically, we selected four actively maintained projects---\texttt{blade}, \texttt{ofdrw}, \texttt{cron-utils}, and \texttt{hutool}---based on their recent activity (commits, accepted pull requests, or issue responses within six months prior to the time of submission).  
To minimize burden on maintainers, we choose the tests to be submitted based on the following criteria:
(1) the focal method must already be covered by at least one existing test, indicating it is an important and stable API; and
(2) when a focal method had more than two existing tests, our generalized tests typically included them, so only the additional ones were submitted.  
Moreover, we manually adjust the tests only for project conventions, including removing or adding comments and renaming variables (e.g., changing \texttt{plainPasswd} to \texttt{plain} for consistency with existing tests).

\autoref{tab:field_study} reports the results of the field study, including the number of accepted, pending, and rejected submitted tests.
In total, we submitted 5 pull requests containing 27 tests for 10 focal methods.
Among them, 16 tests have been accepted, 11 tests remain pending, and none are rejected.
As an example, in the motivating case of \texttt{setPaint(Paint paint)} from project \texttt{ofdrw} (Section~\ref{sec:motivation}), we submitted two tests: \texttt{setPaintGradientPaint()} and \texttt{setPaintTexturePaint()}.  
The former has been accepted, while the latter remains pending.
Because the project currently does not provide complete functionality for texture paint,
the pending test may have uncovered a new requirement, which may be under consideration by the maintainers.
\rev{We also evaluated ChatTester on these focal methods.
ChatTester failed to generate 37.0\% (10/27) of the generalized tests.
Importantly, among the 16 tests accepted by developers, ChatTester missed 37.5\% (6/16).
This confirms that \tool produces valuable tests SOTA approaches miss.}

\section{Discussion}
\subsection{Limitations and Future Work}
\subsubsection{Quality of Generated Oracles}
Oracle generation has long been an open problem in test generation.
The quality of oracles generated by \tool is potentially affected by bugs in the projects under test.
In our evaluation, this issue is minimized because the dataset consists of well-established projects where focal methods are already equipped with multiple developer-written tests.  
However, in practice, developers cannot always assume correctness. 
If a focal method or related project knowledge contains bugs, then the oracles deduced from them may also be incorrect.

To mitigate this, \tool (as described in Section~\ref{sec:test_generalization}) generates not only a primary oracle---derived from the focal method implementation and collected knowledge---but also alternative oracles inferred from the common knowledge embedded in LLMs.  
Developers can intervene in the generalization process and confirm or refine the intended oracle when alternatives are presented.  
In a preliminary experiment, we found that \tool has a 72.98\% probability of generating at least one alternative oracle capable of revealing the same synthetic bugs as the ground-truth tests.  
Future work will further explore strategies to improve the quality and reliability of generated oracles.

\subsubsection{Completeness of Knowledge Collection}
\tool prompts the LLM to proactively query crucial knowledge based on its analysis and reasoning.
In Stage 1, knowledge is collected via examination: the LLM answers multiple-choice exams to determine whether it requires relevant project knowledge to correctly understand the test scenarios.  
A limitation of this design is that the LLM may guess the correct oracle, causing important knowledge to be missed.  
Future work will investigate more robust strategies, such as exploring the project’s knowledge graph, to improve the completeness of knowledge collection and reduce reliance on chance.

\subsection{Threats to Validity}
\noindent \textbf{Language generalizability.}
\rev{Evaluated only on Java, the transferability of auto-tuned rules to other languages remains uncertain.}
However, \tool's pipeline is language-agnostic, and the involved tools like \texttt{CodeQL} and \texttt{LSP} naturally support multiple programming languages. 
\rev{Adapting to a new language mainly requires re-running prompt auto-tuning on a target-language dataset.}

\noindent \textbf{LLM-assessed scenario coverage.}
The nondeterministic nature of LLMs may affect the reliability of the LLM-assessed scenario coverage metric.
To alleviate this, we set the temperature to zero, constraining randomness and ensuring more deterministic responses.
We also repeat the evaluation three times to confirm its reliability and report the averaged results.

\noindent \textbf{Emphasis on developer-written tests.}
Our primary metric (i.e., scenario coverage) measures how well generalized tests cover the scenarios targeted by developer-written ground-truth tests. 
Generalized tests that go beyond the ground truth may capture additional valuable scenarios and should not be deemed redundant by default.
Assessing the value of such extra scenarios requires feedback from project maintainers, making a fully automated, large-scale evaluation infeasible.
To mitigate this, we conducted a field study, which demonstrated the practical value of such tests.

\noindent \textbf{Initial-test assumption.}
\tool assumes an initial test as input for generalization, representing the first test a developer writes for a focal method. 
\rev{This makes \tool less fully automated than end-to-end test generation approaches. 
However, our goal is to augment developer workflows by generalizing an existing test into broader intention-aligned scenarios, rather than replacing fully automated tools. 
An initial test could also be generated by automated approaches,
making \tool naturally composable with existing test generation techniques.}

\section{Related Work}
\label{sec:related-work}

\textbf{Coverage-driven software testing.}
Software testing has traditionally been treated as a constraint-solving problem, where the goal is to generate tests that cover targeted program branches.  
Representative approaches include symbolic execution (both dynamic and static)~\cite{cadar2008klee, braione2017combining, braione2018sushi, godefroid2005dart, sen2005cute} and search-based testing~\cite{fraser2011evosuite, arcuri2008search, braione2017combining, godoy2021enabledness, lin2021graph, lemieux2023codamosa, pacheco2007randoop, lin2020recovering}.  
Although coverage-driven techniques can, to some extent, address test generalization, they cannot capture requirements or infer test scenarios, which do not necessarily correspond to control-flow branches.  
This limitation motivates our design of \tool, which leverages LLMs for requirement understanding and scenario generalization.

\noindent \textbf{LLM-based test generation.}
With the rise of LLMs, 
recent work applies LLMs to software testing~\cite{li2023nuances, schafer2023empirical, xia2024fuzz4all, alshahwan2024automated, generator_yanjie, gao2025promptalchemistautomatedllmtailored, mezzaro2024empirical,lessismore,test_rel_1,wen2025variable,issta24_test_adaption,qi2025intention,qi2021dreamloc,wang2025rulepilotllmpoweredagentsecurity}.
RulePilot~\cite{wang2025rulepilotllmpoweredagentsecurity} leverages an intermediate representation (IR) to structurally capture security semantics, enabling the systematic derivation of both detection rules and corresponding test cases with improved consistency and scenario coverage.
One closely related work is ChatTester~\cite{chattester}, which generates a single test for a given focal method.  
In contrast, \tool generalizes from an initial test to produce multiple tests that comprehensively cover developer-intended scenarios.  
Another relevant work is IntUT~\cite{IntUT}, which first generates test intentions for a focal method---each corresponding to a specific branch and specifying input parameters and expected outputs---and then generates test cases from these intentions to maximize coverage.  
Nevertheless, IntUT is essentially a coverage-driven approach that focuses on branch coverage.  
By contrast, \tool targets scenario coverage, which is requirement-driven rather than code-coverage-driven.

{
\color{black}
\noindent \textbf{Property-based test generation.}
Property-based testing~\cite{maciver2019hypothesis,vikram2023can} and parameterized unit tests~\cite{proze,tiwari2021production} generalize inputs against developer-specified invariants, whereas \tool infers behavioral intent directly from an example test without requiring explicit properties.
For example, PROZE~\cite{proze} derives parameterized tests from runtime data, primarily generalizing observed input values. In contrast, \tool performs scenario-level generalization, identifying variation points that affect both inputs and oracle logic without relying on execution traces.
}

\section{Conclusion}
This work presents \tool, a framework that generalizes developer-written tests to cover requirement-driven scenarios beyond traditional code coverage–driven approaches. 
Given a focal method with an initial test,
\tool generates scenario templates, instantiates them into concrete scenarios, and produces diverse and practical tests. 
Evaluation on 506 focal methods and 1,637 scenarios demonstrates that \tool significantly outperforms state-of-the-art baselines.
A field study further demonstrates its practical value.

\section{Data Availability}
All source code, benchmark, experimental results, and field study data are available at \url{https://github.com/code-philia/TestGeneralizer}.

\begin{acks}
This research is supported in part by the National Natural Science Foundation of China (62572300), the Minister of Education, Singapore (MOE-T2EP20124-0017, MOET32020-0004), the National Research Foundation, Singapore and the Cyber Security Agency under its National Cybersecurity R\&D Programme (NCRP25-P04-TAICeN), DSO National Laboratories under the AI Singapore Programme (AISG Award No: AISG2-GC-2023-008-1B), and Cyber Security Agency of Singapore under its National Cybersecurity R\&D Programme and CyberSG R\&D Cyber Research Programme Office. Any opinions, findings and conclusions or recommendations expressed in this material are those of the author(s) and do not reflect the views of National Research Foundation, Singapore, Cyber Security Agency of Singapore as well as CyberSG R\&D Programme Office, Singapore.
\end{acks}

\bibliographystyle{ACM-Reference-Format}
\bibliography{references}

\end{document}